\documentclass[%
 reprint,
 pra,
 amsmath,amssymb,
 aps,
]{revtex4-2}
\usepackage{graphicx}
\usepackage{dcolumn}
\usepackage[T1]{fontenc}
\usepackage[utf8]{inputenc}
\usepackage{amsmath}
\usepackage{physics}
\usepackage[export]{adjustbox}
\usepackage{float}
\usepackage{indentfirst}
\usepackage{bm}
\usepackage{xcolor}
\usepackage[normalem]{ulem} 
\usepackage[colorlinks=true, allcolors=blue,bookmarks=true]{hyperref} 
\usepackage{cleveref}
\usepackage{tikz}
\usetikzlibrary{quantikz}
\usepackage{tikzit}
\usepackage{natbib} 
\usepackage{subcaption}
\usetikzlibrary{external}
\usepackage[english]{babel}

\date{April 2022}

\begin{document}

\preprint{APS/123-QED}

\title{Implementation of a two-stroke quantum heat engine with a collisional model}

\author{Filipe V. Melo, Nahum Sá, Itzhak Roditi, Alexandre M. Souza, Ivan S. Oliveira, Roberto S. Sarthour}
\affiliation{Centro Brasileiro de Pesquisas Físicas, Rio de Janeiro, Brazil}
\author{Gabriel T. Landi}
\affiliation{Instituto de Física da Universidade de São Paulo, São Paulo, Brazil}
\affiliation{School of Physics, Trinity College Dublin, College Green, Dublin 2, Ireland}

\begin{abstract}
     We put forth a quantum simulation of a stroboscopic two-stroke thermal engine in the IBMQ processor. The system consists of a quantum spin chain connected to two baths at their boundaries, prepared at different temperatures using the variational quantum thermalizer algorithm.
     The dynamics alternates between heat and work strokes, which can be separately designed using independent quantum circuits. 
     The results show good agreement with theoretical predictions, showcasing IBMQ as a powerful tool to study thermodynamics in the quantum regime, as well as the implementation of variational quantum algorithms in real-world quantum computers. 
     It also opens the possibility of simulating quantum heat transport across a broad range of chain geometries and interactions.
\end{abstract}

\maketitle

\section{Introduction}\label{sec:introduction}

Energetics of quantum devices is an active topic of research~\cite{Auffeves_2021}, with many unique features.
Quantum chains may present anomalous heat transport~\cite{Bertini_2021,Landi_2021}, negative differential conductivity~\cite{Mendoza_Arenas_2013}, and perfect rectification~\cite{Balachandran_2018}.
Often, the interaction energy between two systems is comparable to that of their individual parts~\cite{Jarzynski2017,Perarnau2018,Strasberg2019,Talkner2020}, causing the local notion of energy, as belonging to individual systems, to break down. 
As a consequence, the separation between heat and work may  become non-trivial, specially in the presence of quantum coherence. 
In the quantum domain, other resources also come into play: Quantum correlations, for example, can be consumed to make heat flow from cold to hot~\cite{Partovi2008,Jennings2010,Micadei2017}, similar to how electric energy is consumed to run a fridge. 
Finally, the invasive nature of quantum measurements makes all the above quantities extrinsic to the specific choice of measurement protocol~\cite{Perarnau2017,Micadei2020,Levy2020,Micadei2021}. 

The above arguments highlight the need for further experiments, able to assess the energetics of specific quantum devices. 
In this respect, quantum heat engines~\cite{Kosloff_2014,Mitchison_2019} are particularly interesting, as they epitomize the fundamental questions of the field. 
Several experimental demonstrations of quantum heat engines have been put forth in recent years, including in 
trapped ions~\cite{Ro_nagel_2016,von_Lindenfels_2019,Maslennikov_2019,Van_Horne_2020}, single electron boxes~\cite{Koski_2014,Koski_2015}, superconducting devices on the IBMQ network~\cite{Solfanelli_2021}, nuclear magnetic resonance~\cite{Peterson_2019,Denzler_2021},
and 
nitrogen vacancy centers~\cite{Klatzow_2019}. 

In this paper we provide an experimental quantum simulation of a stroboscopic, two-stroke quantum heat engine in the IBM Quantum processor~\cite{ibmq}, based on a collisional model.
We focus on two-stroke engines, which alternate between heat and work strokes, in a generalization of the so-called SWAP engines~\cite{Scarani_2002,Quan_2007,Allahverdyan_2010,Uzdin_2014,Campisi_2014,swap_engine_campisi_2015,Molitor_2020}.
The working fluid is a one-dimensional quantum chain, with both ends connected to thermal baths at different temperatures.
Implementing thermal baths in quantum processors is notoriously difficult. 
In our setup this is overcome using a variational quantum thermalizer algorithm, as put forth by Verdon et. al in~\cite{verdon2019VQT}, and detailed further below.
For the heat strokes, the sites are uncoupled from each other, and allowed to interact with the baths at the boundaries. 
Conversely, in the work stroke the baths are uncoupled, and the qubits are allowed to interact with their nearest neighbors according to an arbitrary interaction~\cite{Molitor_2020}.
The experiments are performed in the engine, refrigerator and accelerator configurations. 
We study both the transient dynamics, as well as the limit cycle. 

The paper is organized as follows: \cref{sec:basic_idea} provides a theoretical background for the quantum heat engine is provided. The experimental setup is described in \cref{sec:exp_setup}, and the results are discussed in \cref{sec:results}.

\section{Two-Stroke Quantum Heat Engine}\label{sec:basic_idea}

\begin{figure*}
    \begin{subfigure}[b]{0.6\textwidth}\centering
        \includegraphics[width=\columnwidth]{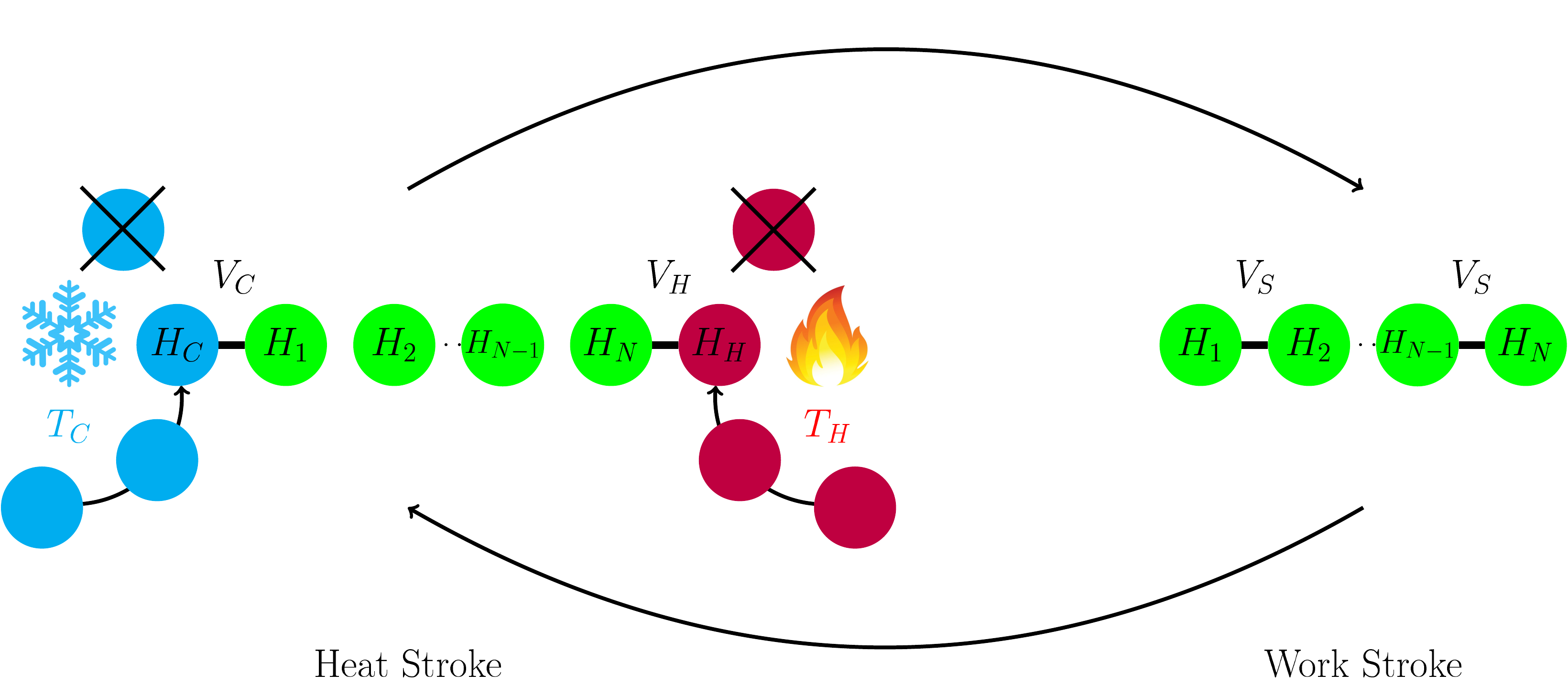}
        \caption{}
    \label{fig:basic_idea}
    \end{subfigure}
    \hspace{5mm}
    \begin{subfigure}[b]{0.35\textwidth}\centering
    \includegraphics[width=\columnwidth]{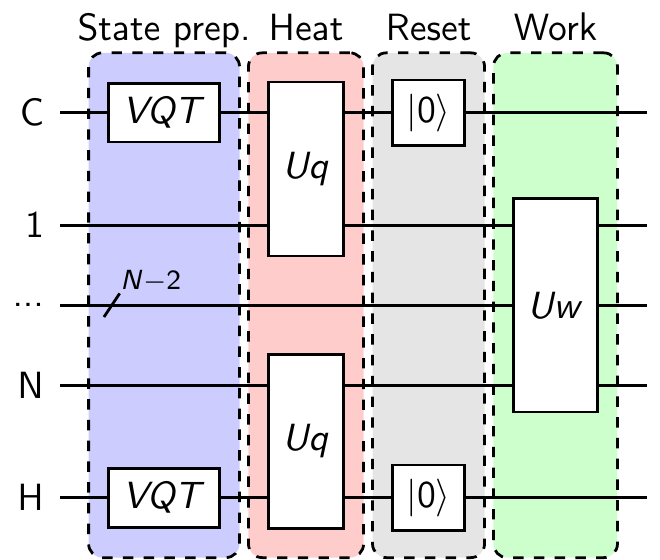}
    \caption{}
    \label{fig:exp_scheme}
    \end{subfigure}
    \caption{\small{The quantum heat engine operates in two modes, as shown in (\subref{fig:basic_idea}): The heat stroke, and the work stroke. In the heat stroke, the internal interactions of the quantum chain are turned off and the system boundary qubits interact with a cold bath at temperature $T_C$ and a hot bath at temperature $T_H$, both described by auxiliary qubits. In the work stroke, the baths are uncoupled and the chain qubits interact with their nearest neighbors. The strokes are operated in a cyclic way, and before each heat stroke the auxiliary qubits are discarded and new ones are introduced, according to the collisional model. The one-cycle circuit configuration, shown in (\subref{fig:exp_scheme}), consists of $N+2$ qubits (in our case $N = 2$). At the beginning, bath qubits $C$ and $H$ in the $|0\rangle$ state are processed with the circuit found with the variational quantum thermalizer (VQT) algorithm to prepare them in thermal states at temperatures $T_C$ and $T_H$. A set of gates implement the heat stroke via a unitary $U_q$, when qubits $1$ and $N$ are measured. The experiment is repeated until the work stroke, which is implemented via an unitary $U_w$, after which all the qubits in the chain are measured. The auxiliary qubits are reset to $|0\rangle$ for the next cycle.}}
    \label{fig:basic_idea_and_exp_scheme}
\end{figure*}

The basic idea is depicted in \cref{fig:basic_idea}. Our design, shown in \cref{fig:exp_scheme}, involves two separate and independent circuit implementations for the heat and work strokes. 
This has the unique advantage that it holds for any internal interaction during the work stroke.
As a consequence, it can be implemented for any chain size and interaction that is programmable on the simulator. 

The two-stroke heat engine we implement consists of a quantum chain with $N$ uncoupled qubits, each with local Hamiltonian $H_i$, and prepared in an arbitrary global state $\rho_S$. 
The qubits can interact with their nearest neighbors, with a Hamiltonian  $V_s = \sum_i V_{i,i+1}$ that can be turned on and off.
In addition, one can also turn on and off an interaction between the qubits at the boundaries and two heat baths, modeled by identical and independently prepared qubits, with Hamiltonians $H_x$, and prepared in thermal states $\rho_x = e^{-H_x/T_x}/Z_x$, with $x \in \{C, H \}$, being $C$ for cold and $H$ for hot, $Z_x$ is the partition function, and $T_C < T_H$ (we set $k_B = 1$).

The dynamics alternate between heat and work strokes. During the heat stroke, the internal interactions are turned off, while the boundary qubits interact with their respective baths $C$ and $H$ via Hamiltonians $V_{C}$ and $V_{H}$. The system thus evolves for a period $\tau_q$, under the action of a Hamiltonian, $H_q = \sum_i H_i + H_C + H_H + V_{C} + V_{H}$, and with an evolution characterized by the map
\begin{equation}
    \Tilde{\rho}_S = \mathcal{E}_q(\rho_S) = \Tr_{CH}\{ U_q(\rho_C\rho_S\rho_H)U_q^{\dag} \},
\end{equation}
where $U_q = e^{-iH_q \tau_q}$.
In the work stroke, the baths are disconnected from the chain and the internal interactions $V_S$ are turned on, allowing energy to flow through the chain. The system evolves for a time $\tau_w$ under the action of the Hamiltonian $H_w = \sum_i H_i + V_S$, according to the map
\begin{equation}
    \rho'_S = \mathcal{E}_w(\rho_S) = U_w \Tilde{\rho}_S U^{\dag}_w,
\end{equation}
where $U_w = e^{-i H_w \tau_w}$. The execution of the two strokes in succession results in a cycle of duration $\tau = \tau_q + \tau_w$. 
Fig.~\ref{fig:exp_scheme} illustrates one such cycle. 
Crucially, for each new cycle the baths $C$ and $H$ are reset to  the same temperatures $T_C$ and $T_H$, as in a collisional model~\cite{de2018reconciliation}. 
The machine thus follows a stroboscopic dynamics.
Letting $\rho^n_S$ denote the initial state of the $n^{th}$ cycle, and $\Tilde{\rho}^n_S$ the intermediate state between the strokes, the dynamics is given by
\begin{equation}
    \Tilde{\rho}_S^n = \mathcal{E}_q(\rho^n_S), \qquad
    \rho_S^{n+1} = \mathcal{E}_w(\Tilde{\rho}^n_S).
\end{equation}
By splitting the dynamics in two strokes, we have reduced the problem to two independent simulation designs. For the work stroke, one need only to design $U_w$. Conversely, for the heat stroke, one must design both $U_q$ and the thermal states of the baths.
The implementation of the latter is in practice quite difficult, and is overcome here using the variation quantum thermalizer algorithm~\cite{verdon2019VQT}, as explained in \cref{sec:variational_quantum_thermalizer}.

In order to track the energetics of the engine, we can measure the energy changes of the baths and the system during each cycle. 
In the heat stroke of the $n^{th}$ cycle, we define heat as the change in energy of the baths:
\begin{equation}
    Q^n_x = -\Tr{H_x(\Tilde{\rho}^n_x - \rho_x)} \hspace{40pt}
\end{equation}
which are positive when energy enters the system.
Notice that $\rho_x$ is the same for all cycles, since the baths are reset. 
The heat $Q^n_x$ is, in general, different from the change in energy of the system, as there may be a work cost associated to turning on/off the interactions $V_{C(H)}$. From global energy conservation, we find that~\cite{de2018reconciliation}:
\begin{equation}
\begin{split}
    W^{\text{on/off}}_C & = Q^n_C + \Tr{H_1(\Tilde{\rho}^n_S - \rho^n_S)} = -\Delta V^n_C \\
    W^{\text{on/off}}_H & = Q^n_H + \Tr{H_N(\Tilde{\rho}^n_S - \rho_S^n)} = -\Delta V^n_H
\end{split}
\end{equation}
with $\Delta V^n_x = \Tr{V_x(\Tilde{\rho}^n_{CSH}-\rho_{C}\rho^n_{S}\rho_{H})}$. 

This work term only vanishes when 
\begin{equation}\label{eq:str_en_cons}
    [V_C, H_1 + H_C] = [V_H, H_N + H_H] = 0,
\end{equation}
a condition known as \textit{strict energy conservation} \cite{Molitor_2020}. When this is satisfied, we can write
\begin{equation}\label{eq:heats}
\begin{split}
    Q^n_C & = \Tr{H_1(\Tilde{\rho}^n_S - \rho^n_S)} \\ 
    Q^n_H & = \Tr{H_N(\Tilde{\rho}^n_S - \rho^n_S)}
\end{split}
\end{equation}
meaning that all energy that leaves the bath enters the system and vice-versa. Thus, one can determine the change in energy during the heat stroke measuring only the state of the system chain. 

During the work stroke, the interactions $V_S=\sum_i V_{i,i+1}$ are turned on so that energy is allowed to flow through the chain. The work cost associated to this is now connected to turning $V_S$ on and off:
\begin{equation}\label{eq:work}
\begin{split}
    W^n &= -\Tr{\Big(\sum_i H_i \Big)(\rho^{n+1}_S - \Tilde{\rho}^n_S)}\\ 
    &= \Tr{V_S(\rho^{n+1}_S - \Tilde{\rho}^n_S)}
\end{split}
\end{equation}
which is defined as positive when energy leaves the system.

After several strokes, the engine will reach a limit cycle, such that
\begin{equation}
    \Tilde{\rho}_S^* = \mathcal{E}_q(\rho^*_S), \hspace{30pt}
    \rho_S^* = \mathcal{E}_w(\Tilde{\rho}^*_S).
\end{equation}
The system therefore alternates between  $\Tilde{\rho}_S^*$ and $\rho_S^*$ after each stroke, and \cref{eq:heats,eq:work} become
\begin{align}
    Q^*_C &= \Tr{H_1(\Tilde{\rho}^*_S - \rho^*_S)} \\
    Q^*_H &= \Tr{H_N(\Tilde{\rho}^*_S - \rho^*_S)} \\
    W^* &= -\Tr{\Big(\sum_i H_i \Big)(\rho^*_S - \Tilde{\rho}^*_S)}
\end{align}
At the limit cycle, the system internal energy no longer changes so that 
\begin{equation}\label{eq:first_law}
    Q^*_C + Q^*_H = W^*,
\end{equation}
which is the first law. 

\section{Experimental Setup}\label{sec:exp_setup}

In this paper, we present an implementation of the just-described engine
in \textit{ibm\_lagos} (v1.0.8), which is one of the IBM Quantum Falcon processors, based on superconducting qubits \cite{krantz2019quantum}. The use of this quantum processor is accessible through the cloud along with other similar devices in \cite{ibmq}, and can be remotely controlled via a personal computer with the open-source Python framework \textit{Qiskit} \cite{Qiskit}, also offered by IBM.

\subsection{System Configuration}\label{sec:system_configuration}

The system consists of a chain of $N=2$ non-resonant qubits ($\omega_1 \neq \omega_2$), with Hamiltonians:
\begin{equation}\label{eq:qchain_hamilt}
    H_1 = \frac{\omega_1}{2}\sigma^1_z,\hspace{30pt} H_2 = \frac{\omega_2}{2}\sigma^2_z,
\end{equation}
and two ancillary qubits to represent the thermal baths:
\begin{equation}\label{eq:baths_hamilt}
    H_C = \frac{\omega_C}{2}\sigma^C_z, \hspace{30pt} H_H = \frac{\omega_H}{2}\sigma^H_z.
\end{equation}
All the interactions are chosen to be of the form
\begin{equation}\label{eq:interactions}
    V_{j,k} = g_{j,k}\big( \sigma_+^{j}\sigma_-^{k} + \sigma_-^{j}\sigma_+^{k} \big)
\end{equation}
where $\sigma_{\pm} = \sigma_x \pm i \sigma_y$. This concerns the heat stroke interactions $V_C = V_{1,C}$ and $V_H = V_{2,H}$, as well as the work stroke interaction $V_{1,2}$. The baths are assumed to be resonant with their sites, ($\omega_C = \omega_1$ and $\omega_H = \omega_2$), from which one finds that \cref{eq:interactions} satisfies the conditions (\ref{eq:str_en_cons}).

From \cref{eq:heats}, the heats for the $n^{th}$ cycle are given by
\begin{equation}
\begin{split}
    Q^n_C &= \frac{\omega_1}{2}\big(\expval{\Tilde{\sigma}_z^1}_n-\expval{\sigma_z^1}_n \big)\\
    Q^n_H &= \frac{\omega_2}{2}\big(\expval{\Tilde{\sigma}_z^2}_n-\expval{\sigma_z^2}_n \big)
\end{split}
\end{equation}
where $\expval{\sigma_z^i}_n = \Tr{\sigma_z^i \rho_S^n}$ and 
$\expval{\Tilde{\sigma}_z^i}_n = \Tr{\sigma_z^i \Tilde{\rho}_S^n}$. 
Following \cref{eq:work}, the work for the $n^{th}$ cycle is
\begin{equation}
    W^n = -\sum_{i=1,2}\frac{\omega_i}{2}\big(\expval{\sigma_z^i}_{n+1}-\expval{\Tilde{\sigma}_z^i}_n \big).
\end{equation}

\subsection{Hardware}\label{sec:hardware}

\begin{figure}[t]
    \includegraphics[width=0.4\columnwidth]{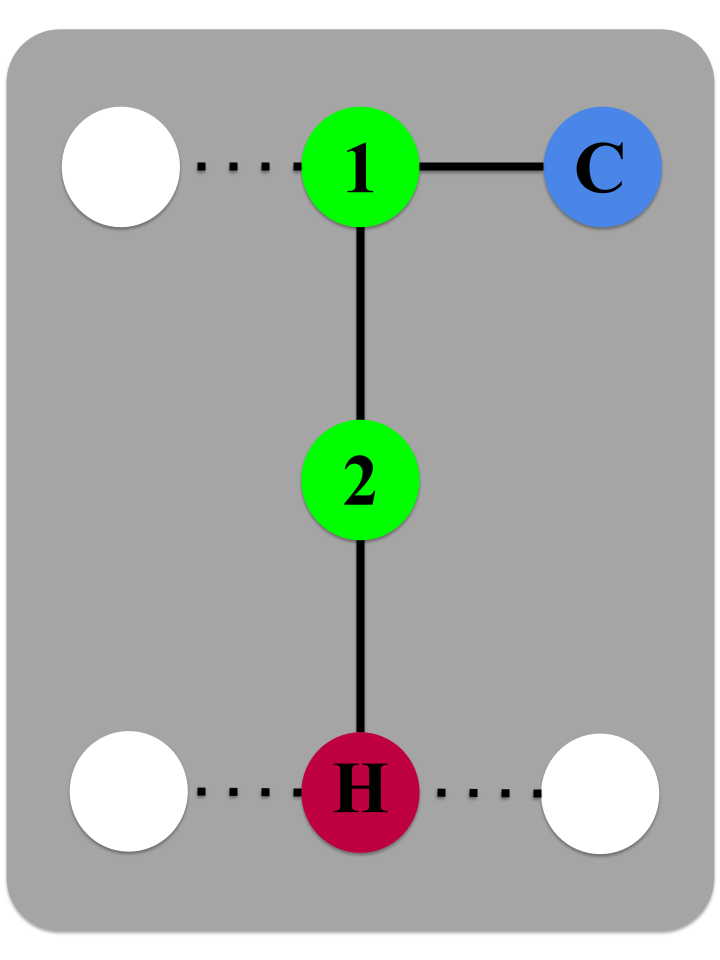}
    \caption{The picture represents the topology of the quantum chip \textit{ibm\_lagos}. Nodes and edges represent qubits and their connections, respectively. Green nodes represent the quantum chain with two qubits. Blue and red nodes represent the auxiliary qubits that are used as the cold and hot baths, respectively. White nodes and dashed edges are qubits and connections that were not used during the experiment.}
    \label{fig:processor_config}
\end{figure}

The topology of the 7-qubit \textit{ibm\_lagos} processor is shown in \cref{fig:processor_config}. This superconducting device can apply z-rotations, $X$ and $\sqrt{X}$ gates, reset operations and CNOT gates between connected qubits. These gates are implemented as microwave pulses that rotate the qubits in the Bloch sphere. The properties of the qubits, gates and connections of the processor are provided in \cref{apx:system_properties}. Each qubit is initially prepared in the $|0\rangle$ state, and by measuring it one have its projection on the computational basis. Thus, a quantum circuit must be executed several times in order to build a statistics of the measurements, each repetition called a shot. The number of shots can be defined by the user before running an experiment.

When mapping a quantum circuit to an IBM Quantum processor, any gate that is not in this set must be expressed in terms of the available operations. Also, the connectivity of the topology must be considered, since applying CNOT gates to unconnected qubits requires SWAP operations, which are expensive to perform on a noisy quantum device. All these steps may be performed automatically with Qiskit by a process called \textit{transpilation} \cite{Qiskit, retworkx}. Besides rewriting a quantum circuit to match the topology of the real device, the transpilation also optimizes it for execution on noisy systems. In our problem, the interactions are all between neighboring qubits, thus, a good choice of layout is that shown in \cref{fig:processor_config}.

\subsection{Quantum Simulation}\label{sec:quantum_simulation}

The dynamics of the system is implemented via quantum simulation, a process which consists of reproducing the dynamics of a quantum system using another quantum system \cite{feynman1982simulating}. For a system in an initial state $|\psi(0)\rangle$ under the action of a time-independent Hamiltonian $H_S$, the time evolution is given by the solution of the Schrödinger equation
\begin{equation}\label{eqn:evolution_schrodinger}
    |\psi (t)\rangle = e^{-iH_S t/\hbar}|\psi(0)\rangle,
\end{equation}
The problem consists of simulating the dynamics of $H_S$ given by (\ref{eqn:evolution_schrodinger}), using operations that can be implemented in the physical system that will act as the quantum simulator.

\begin{figure}[b]
    \centering
    \includegraphics[width=\linewidth]{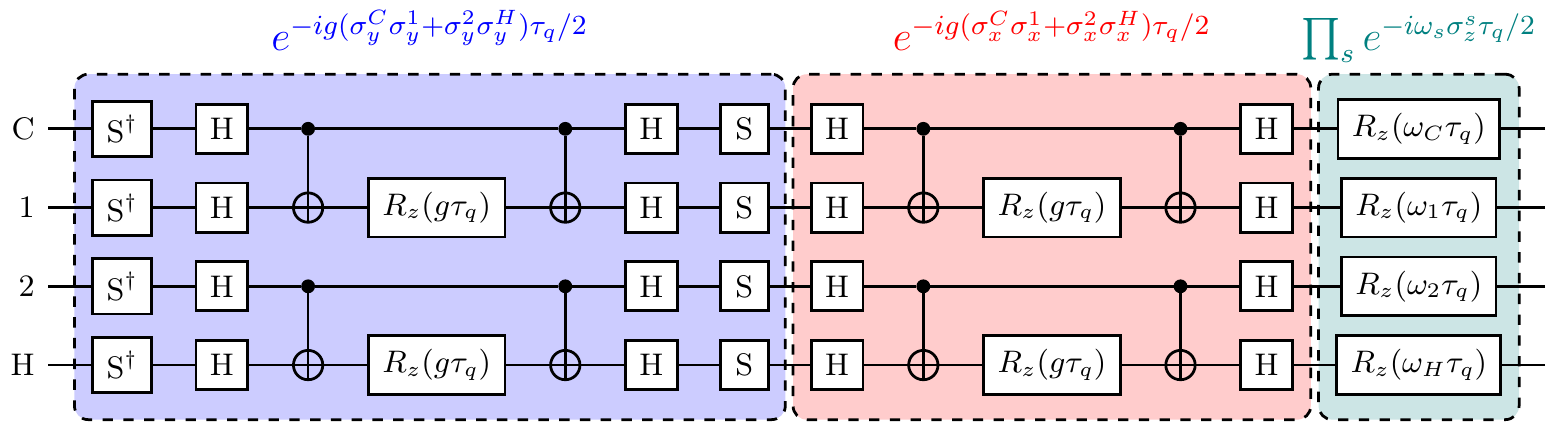}
    \includegraphics[width=\linewidth]{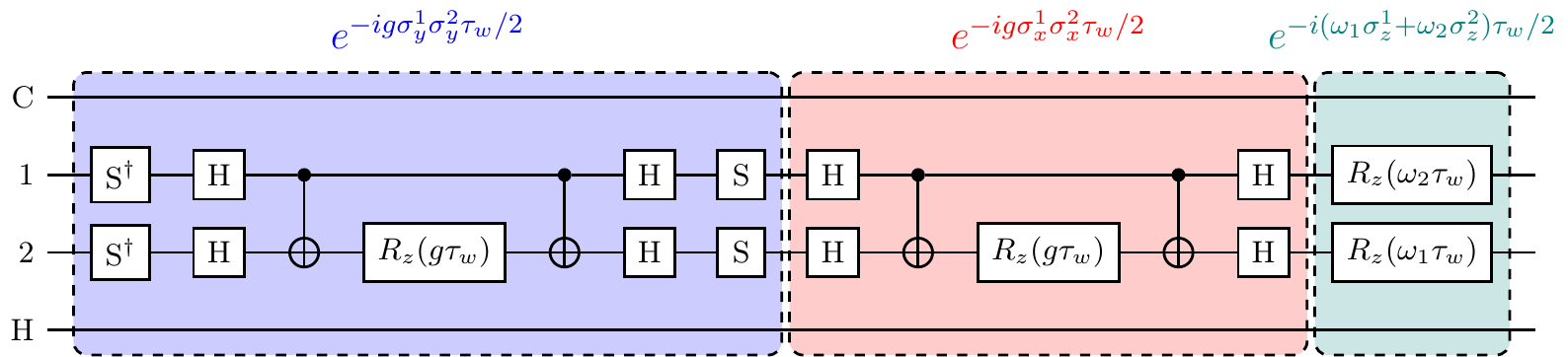}
    \caption{Quantum circuits for the implementation of the heat (above) and work (below) strokes. S and H represent phase and Hadamard gates, respectively, and $R_z(\theta)$ are rotations of an angle $\theta$ around the $z$-axis. The product on the green part of the first circuit is performed on the set $s=\{1,2,C,H\}$.}
    \label{fig:stroke_circs}
\end{figure}

The evolution operators $U_q$ and $U_w$ for the heat and work strokes may be conveniently written as
\begin{equation}\label{eq:uq_parts}
\begin{split}
    U_q = & e^{-i g (\sigma^C_y \sigma^1_y + \sigma^2_y \sigma^H_y) \tau_q / 2} \times \\
    & e^{-i g (\sigma^C_x \sigma^1_x + \sigma^2_x \sigma^H_x) \tau_q / 2} \times \\
    & {e^{-i (\omega_1 \sigma^1_z + \omega_2 \sigma^2_z + \omega_C \sigma^C_z + \omega_H \sigma^H_z) \tau_q / 2}}
\end{split}
\end{equation}
\begin{equation}\label{eq:uw_parts}
\begin{split}
    U_w = & e^{-i g (\sigma^C_y \sigma^1_y + \sigma^2_y \sigma^H_y) \tau_w / 2} \times \\
    & e^{-i g (\sigma^C_x\sigma^1_x + \sigma^2_x \sigma^H_x) \tau_w / 2} \times \\
    & {e^{-i (\omega_1 \sigma^1_z + \omega_2 \sigma^2_z) \tau_w / 2}}
\end{split}
\end{equation}
Thus, one must find a set of single- and two-qubit quantum gates $\mathcal{U}_m$ and $\mathcal{W}_m$ such that
\begin{equation}
        U_q \approx \prod^M_{m=0} \mathcal{U}_m, \hspace{30pt}
        U_w \approx \prod^M_{m=0} \mathcal{W}_m.
\end{equation}
Such set can be found using the \textit{Operator Flow} module from Qiskit, and the resulting quantum circuit is shown in \cref{fig:stroke_circs}, with the parts that simulate each factor of \cref{eq:uq_parts} and \cref{eq:uw_parts} being identified in the picture.

\subsection{Variational Quantum Thermalizer}\label{sec:variational_quantum_thermalizer}

At the beginning of each cycle the thermal states of the baths are prepared using a quantum circuit found with the variational quantum thermalizer algorithm (VQT), which belongs to a class of variational algorithms called Quantum-Hamiltonian Based Models (QHBM).
Given a Hamiltonian $H$ and a target inverse temperature $\beta = 1/T$, the goal of the VQT is to generate the best approximation to the thermal state:
\begin{equation}\label{eq:thermal_state}
    \rho_{\beta} = \frac{1}{\mathcal{Z}_{\beta}}e^{-\beta H}, \hspace{20pt} \mathcal{Z}_{\beta} = \Tr(e^{-\beta H})
\end{equation}
where $\mathcal{Z_{\beta}}$ is the partition function.

We begin with a density matrix $\rho_{\bm{\theta}}$, defined by a set of pure states $\{ |\psi_i\rangle \}$, that constitutes a basis on a Hilbert space of dimension $d$, and a parameter vector $\bm{\theta}$. Being $\{p_i(\theta_i)\}$ the probability distribution corresponding to the \textit{i}-th basis state, such a matrix can be written as 
\begin{equation}\label{eq:mixed_state}
    \rho_{\bm{\theta}} = \sum_{i=1}^{d} p_i(\theta_i)|\psi_i\rangle \langle \psi_i|.
\end{equation}
We sample from this distribution a pure state $|\psi_i\rangle$, pass it through a parameterised quantum circuit $U(\bm{\phi})$, and measure the mean energy  $\expval{H}_i = \expval{U(\bm{\phi}) H U^\dag(\bm{\phi})}{\psi_i}$. Repeating this process many times for the set $\{ |\psi_i \rangle, p_i(\theta_i) \}$ and averaging the obtained $\expval{H}_i$ will give us the mean energy $\expval{H}_{\bm{\theta}\bm{\phi}}$ with respect to $\rho_{\bm{\theta}\bm{\phi}} = U(\bm{\phi})\rho_{\bm{\theta}}U^\dag(\bm{\phi})$:
\begin{equation}\label{eq:mean_energy}
    \expval{H}_{\bm{\theta}\bm{\phi}} = \sum_{i=1}^{d} p_i(\theta_i)\expval{H}_i.
\end{equation}
As in any variational algorithm, in order to find the optimal parameters  $\{ \bm{\theta}', \bm{\phi}' \}$ that solve the problem, we must find the minimum of a \textit{loss function}, which here can be obtained from (\ref{eq:mean_energy}) and the von Neumann entropy $S(\rho_{\bm{\theta} \bm{\phi}})$:
\begin{equation}\label{eq:loss}
    \mathcal{L}(\bm{\theta}, \bm{\phi}) = \beta \expval{H}_{\bm{\theta} \bm{\phi}} - S(\rho_{\bm{\theta}}),
\end{equation}
which is minimized when $\rho_{\bm{\theta} \bm{\phi}}$ is equal to the required thermal state. This loss function would require the measurements of two quantities, but we can take advantage from the fact that the entropy is invariant under unitary transformations on the density matrix: this way only the energy measurement is necessary, as the entropy can be calculated from the probability distribution parameters $\{ \bm{\theta} \}$.

\begin{figure}
    \centering
        COLD BATH
        \includegraphics[width=\linewidth]{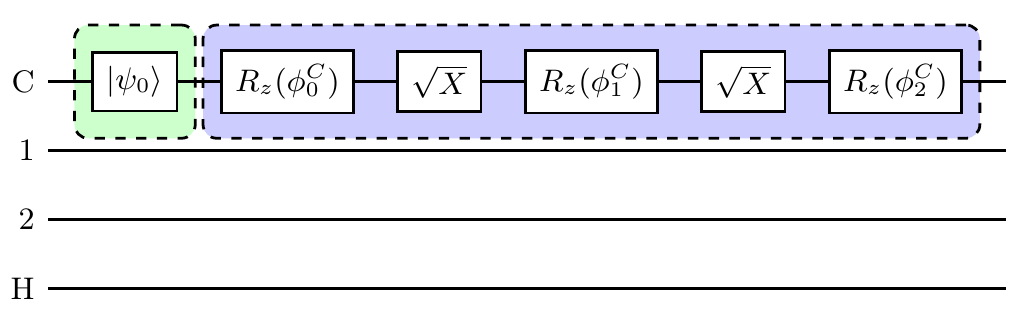}
        \vspace{10pt}
        HOT BATH
        \vspace{10pt}
        \includegraphics[width=\linewidth]{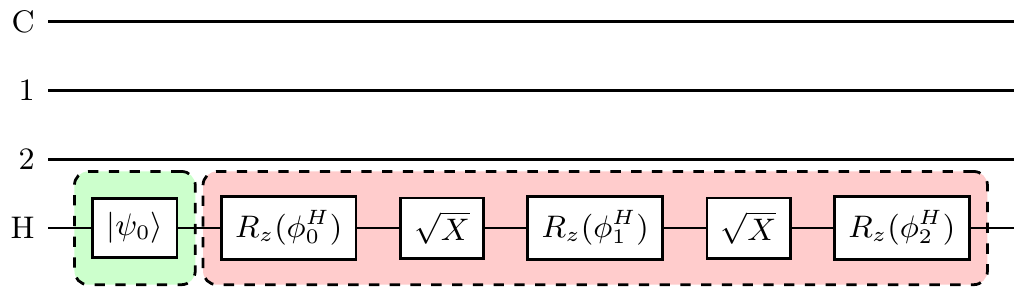}
    \caption{The chosen ansatz for both thermal states consists of three $R_z$ rotations interspersed by square root of NOT gates. Each rotation has a different $\phi$ parameter, which are to be optimized during the VQT implementation. The $|\psi_0 \rangle$ gate sets the initial state to $|0\rangle$ or $|1\rangle$.}
    \label{fig:vqt_ansatzes}
\end{figure}

In our case, the baths have different temperatures $T_C$ and $T_H$, so their thermal state preparation circuits must be found separately with VQT. As they are represented by a single qubit each ($d=2$), one parameter $\theta$ is necessary to describe them, as $\rho_{\theta} = p(\theta)|0\rangle\langle 0| + (1 - p(\theta))|1\rangle\langle1|$, and their parameterized circuits must be executed two times in order to obtain the energy (\ref{eq:mean_energy}). The distribution $p(\theta)$ was chosen to be the sigmoid function $p(\theta) = 1/(1 + e^{-\theta})$. The ansatzes for the cold and hot baths are shown in \cref{fig:vqt_ansatzes}. Their choice is based on the fact that they can approximate any 1 qubit unitary \cite{nielsen2002quantum}, and also take into account the set of gates accessible in \textit{ibm\_lagos} backend, as discussed in \cref{sec:hardware}. Other configurations were also tried, including rotations around other axes, but they did not show optimal results, and some demanded too many iterations of the VQT to converge. Therefore, the loss function $\mathcal{L}(\bm{\theta}, \bm{\phi})$ will contain 4 parameters to be optimized. The result of the VQT for both baths is a factorized state composed of their two subsystems, with dimension $d=4$:
\begin{equation}
    \rho_{CH} = \frac{1}{\mathcal{Z}_{\beta_C}}e^{-\beta_C H_C} \otimes \frac{1}{\mathcal{Z}_{\beta_H}}e^{-\beta_H H_H},
\end{equation}
Thus, the whole experiment must be repeated four times, one for each element of the basis $\{ |0\rangle, |1\rangle \} \otimes \{ |0\rangle, |1\rangle \}$, as we are dealing with mixed states and IBM experiments always start with pure states.

In order to implement the VQT in the experiment, the open-source Python library \textit {Pennylane} \cite{bergholm2020pennylane} was used along with \textit{Qiskit}. The classical optimization was performed using the SciPy library~\cite{2020SciPy-NMeth}, with the \textit{Constrained Optimization BY Linear Approximation} (COBYLA) method \cite{powell1994direct, powell1998direct}, which converged faster than other classical optimizers for our case. In general, it took less than 15 iterations for the algorithm to converge. When running the VQT on a quantum computer, it is important that it is done in the corresponding qubits that will be used as thermal baths, as in this case it finds optimal parameters that consider the device noise associated to those qubits during the learning process. 

\section{Results}\label{sec:results}

\begin{figure*}[t]
    \centering    
    \begin{subfigure}[b]{0.3\textwidth}
        \centering
        Heat Engine
        \includegraphics[width=\linewidth]{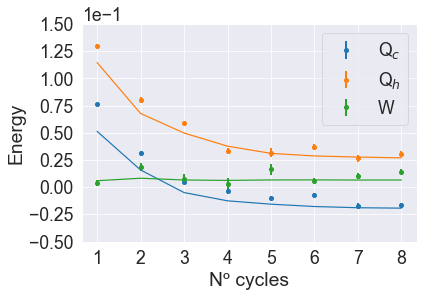}
        \captionsetup{skip=0pt}
        \subcaption{}
        \label{fig:heat_engine}
    \end{subfigure}
    \begin{subfigure}[b]{0.3\textwidth}
        \centering
        Refrigerator
        \includegraphics[width=\linewidth]{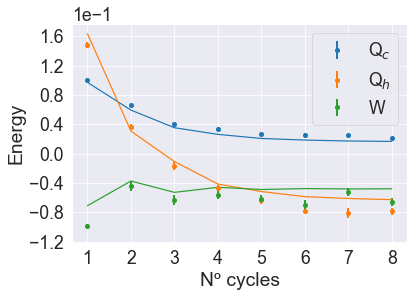}
        \captionsetup{skip=0pt}
        \subcaption{}
        \label{fig:refrigerator}
    \end{subfigure}
    \begin{subfigure}[b]{0.3\textwidth}
        \centering
        Heat Accelerator
        \includegraphics[width=\linewidth]{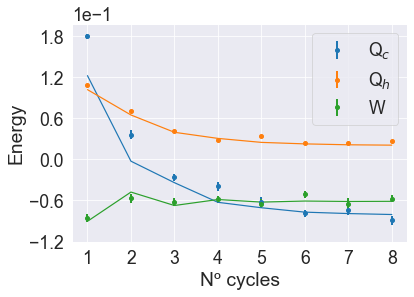}
        \captionsetup{skip=0pt}
        \subcaption{}
        \label{fig:heat_accelerator}
    \end{subfigure}
    
    \begin{subfigure}[b]{0.3\textwidth}
        \includegraphics[width=\linewidth]{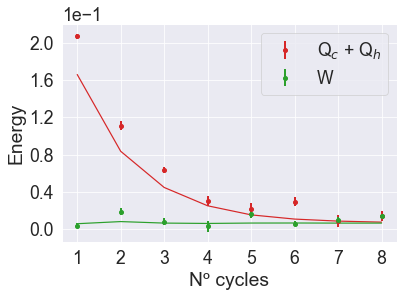}
        \captionsetup{skip=0pt}
        \subcaption{}
        \label{fig:refrigerator_first_law}
    \end{subfigure}
    \begin{subfigure}[b]{0.3\textwidth}
        \includegraphics[width=\linewidth]{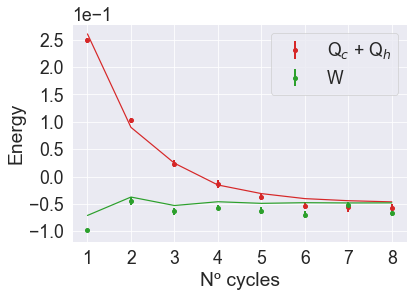}
        \captionsetup{skip=0pt}
        \subcaption{}
        \label{fig:heat_engine_first_law}
    \end{subfigure}
    \begin{subfigure}[b]{0.3\textwidth}
        \includegraphics[width=\linewidth]{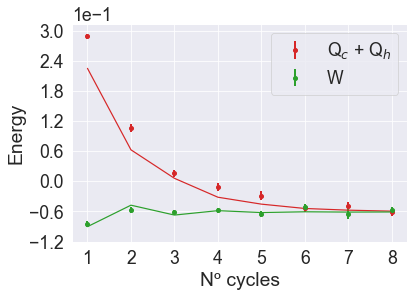}
        \captionsetup{skip=0pt}
        \subcaption{}
        \label{fig:heat_accelerator_first_law}
    \end{subfigure}    
    \caption{\small{Results for the three different modes of operation. First column: heat engine ($\omega_1 = 0.75$, $\omega_2 = 1.0$, $g=0.80$, $T_C = 0.40$, $T_H = 0.80$), second column: refrigerator ($\omega_1 = 0.50$, $\omega_2 = 2.0$, $g=0.80$, $T_C = 1.0$, $T_H = 1.2$), third column: heat accelerator ($\omega_1 = 2.0$, $\omega_2 = 0.50$, $g = 0.80$, $T_C = 1.0$, $T_H = 1.2$). In all cases, both chain qubits were initialized in the ground state. Graphics (a)-(c) depict the dynamics of $Q_C$, $Q_H$ and $W$ during the machine operation; graphics (d)-(f) are a statement of the first law, showing that $Q_C + Q_H$ and $W$ converge to the same value in the limit cycle. The circles represent experimental data, and the solid lines were obtained from a numerical simulation of the Trotter evolution. All results were obtained with 8192 shots (repetitions of the same circuit in order to build a statistics from the counts of $|0\rangle$ and $|1\rangle$). Each experiment was repeated 10 times, and each point represent the mean of the results of the expectation values. The error bars were calculated from three times the standard deviation.}}
    \label{fig:results}
\end{figure*}

The experiments were carried out for three different modes of operation of the machine: heat engine, refrigerator and heat accelerator. They are determined by the values of the frequencies $\omega_1$ and $\omega_2$, and the temperatures $T_C$ and $T_H$ of the baths  \cite{de2018reconciliation,swap_engine_campisi_2015}. These modes can be observed when the machine reaches the limit cycle. In the interval $T_C/T_H \leq \omega_1 / \omega_2 \leq 1$, it operates as a heat engine, withdrawing heat from the hot bath ($Q_H>0$), expelling some of it in the cold bath ($Q_C<0$) and producing some useful work ($W>0$). When $\omega_1 / \omega_2 < T_C/T_H$, it works as a refrigerator, consuming work ($W<0$) in order to extract heat from the cold bath and transfer it to the hot bath ($Q_C>0$, $Q_H<0$). When $\omega_1/\omega_2 > 1$, the machine operates as a heat accelerator, as it consumes work ($W<0$) to accelerate the process of transferring heat from the hot to the cold bath ($Q_H >0$, $Q_C<0$).

The results are shown in \cref{fig:results}. The columns contain the results for each mode of operation. For comparison purposes, a numerical simulation of the Trotter evolution was performed, represented by the solid lines in the graphics. Graphics (a)-(c) shows the dynamics of the energetic flux during the cycles. As expected, the heat engine (a) and accelerator (c) show positive $Q_H$ and negative $Q_C$ in the limit cycle, whereas the refrigerator (b) presents the opposite behavior. During the work stroke, in the limit cycle, the heat engine produced work, as denoted by $W>0$, and the refrigerator and heat accelerator  both consumed work in order to operate, as denoted by $W<0$. Graphics (d)-(f) shows that the system obeys the first law, as stated in \cref{eq:first_law}: in the limit cycle, the sum of the heats $Q_C + Q_H$ and the work $W$ converge to the same quantity.

\section{Conclusions}

In summary, we demonstrated the use of the IBMQ processor as a tool to implement quantum heat engines, with full access to the energetics of the system.
Thermal reservoirs are particularly difficult to implement, which we overcame using 
a variational algorithm at each cycle, effectively implementing a collisional model.
This allowed us to access both the limit cycle, as well as the transient regime, offering unique insights into the relaxation process. 
Our approach cleanly separates the heat and work strokes, and thus can be readily generalized to any other type of heat engines.
This means either chains with multiple qubits, or more complicated interactions, opening up the prospect of simulating quantum heat engines and quantum transport in various many-body systems.

\textit{Acknowledgments.}
This study was financed in part by the Coordenação de Aperfeiçoamento de Pessoal de Nível Superior - Brasil (CAPES) - Finance Code 001, the Brazilian National Institute of Science and Technology for Quantum Information (INCT-IQ) Grant No. 465469/2014-0, the National Council for Scientic and Technological Development (CNPq), and the Carlos Chagas Filho Foundation for Research Support of Rio de Janeiro State (FAPERJ). GTL acknowledges the financial support of the S\~ao Paulo Funding Agency FAPESP (Grant No.~2019/14072-0.), and CNPq (Grant No. INCT-IQ 246569/2014-0). IR also acknowledges CNPq (Grant No. 311876/2021-8). ISO acknowledges support from FAPERJ (Grant No. 202.518/2019). AMS acknowledges support from FAPERJ (Grant No. 203.166/2017). We acknowledge the use of IBM Quantum services for this work. The views expressed are those of the authors, and do not reflect the official policy or position of IBM or the IBM Quantum team. This manuscript used \textit{Qiskit}, \textit{Pennylane} \cite{bergholm2020pennylane} and \textit{QuTiP} \cite{qutip} for the simulations, and \textit{Quantikz} \cite{Quantikz} for drawing the circuits. All code is available on GitHub: \href{https://github.com/lipinor/two-stroke-qhe-vqt}{https://github.com/lipinor/two-stroke-qhe-vqt}.

\appendix

\section{System properties}\label{apx:system_properties}

\begin{table}
    \centering
        \begin{tabular}{|c|c|c|c|c|c|}
            \hline
             Qubit & $T_1$ (us) & $T_2$ (us) & Freq. (GHz) & Meas. err.& Gate err.  \\
             \hline
             C & 142.53 & 101.29 & 5.188 & 6.000E-3 & 1.866E-4 \\
             \hline
             1 & 139.32 & 129.54 & 5.100 & 6.900E-3 & 2.270E-4 \\
             \hline
             2 & 143.70 & 127.84 & 4.987 & 1.850E-2 & 1.238E-4 \\
             \hline
             H & 149.91 & 106.99 & 5.176 & 1.500E-2 & 4.118E-4 \\
             \hline
        \end{tabular}
        \caption{Decoherence times, frequencies and errors for the qubits.}
        \label{tab:backend_config_1}
\end{table}
\begin{table}
        \begin{tabular}{|c|c|c|}
            \hline
              CNOT & Length (ns) & Error \\
             \hline
             C-1 & 327.11 & 5.930E-3 \\
             \hline
             1-2 & 334.22 & 5.081E-3 \\
             \hline
             2-H & 334.22 &  1.053E-2  \\
             \hline
        \end{tabular}
    \caption{Length and associeated errors for the connections.}
    \label{tab:backend_config_2}
\end{table}

Table \ref{tab:backend_config_1} shows properties of the qubits used in the \textit{ibm\_lagos} processor. Average single-qubit gate length is of 35.56 ns. The readout length for all qubits is of 704.0 ns, and its associated error is 0.01. Table \ref{tab:backend_config_2} shows the length of the CNOT gates and their respective errors, for all the connections used in the experiment. Further information about this and other processors can be found in \cite{ibmq}, in the \textit{Services} section.

\bibliographystyle{apsrev4-2}
\bibliography{bibliography.bib}

\end{document}